\documentclass[twocolumn,showpacs,preprintnumbers,amsmath,amssymb,prb]{revtex4-1}
\usepackage{graphicx}
\usepackage{amsmath, amsthm, amssymb}
\usepackage{latexsym}
\usepackage{amssymb}
\usepackage{bm}
\usepackage{dcolumn}
\usepackage{epstopdf}
\usepackage{color}
\usepackage{braket}

\newcommand{\UP}{UPt$_3$}

\begin{document}

\title{Magnetization in the Superconducting State of \UP\ from Polarized Neutron Diffraction }
\date{\today}

\author{W. J. Gannon and W. P. Halperin}
\affiliation{Department of Physics and Astronomy, Northwestern University, Evanston, IL 60660 USA}

\author{C. Rastovski and M. R. Eskildsen}
\affiliation{Department of Physics, University of Notre Dame, Notre Dame, IN 46556 USA}

\author{Pengcheng Dai}
\affiliation{Department of Physics and Astronomy, The University of Tennessee, Knoxville, TN 37996 USA}

\author{A. Stunault}
\affiliation{Institut Max von Laue-Paul Langevin, 38042 Grenoble, France}

\begin{abstract}

The heavy-fermion superconductor \UP\ is thought to have odd parity, a state for which  the temperature dependence of the spin susceptibility is an important signature.  In order to address conflicting reports from two different experiments, the NMR Knight shift and measurements of the anisotropy of the upper critical field,  we have measured the bulk susceptibility in a high-quality single crystal using polarized neutron diffraction. A temperature-independent susceptibility was observed for $H||a$ through the transitions  between the normal state and the superconducting A, B, and C phases, consistent with odd-parity, spin-triplet superconductivity.

\end{abstract}

\maketitle

\begin{figure}
\includegraphics[width=79mm]{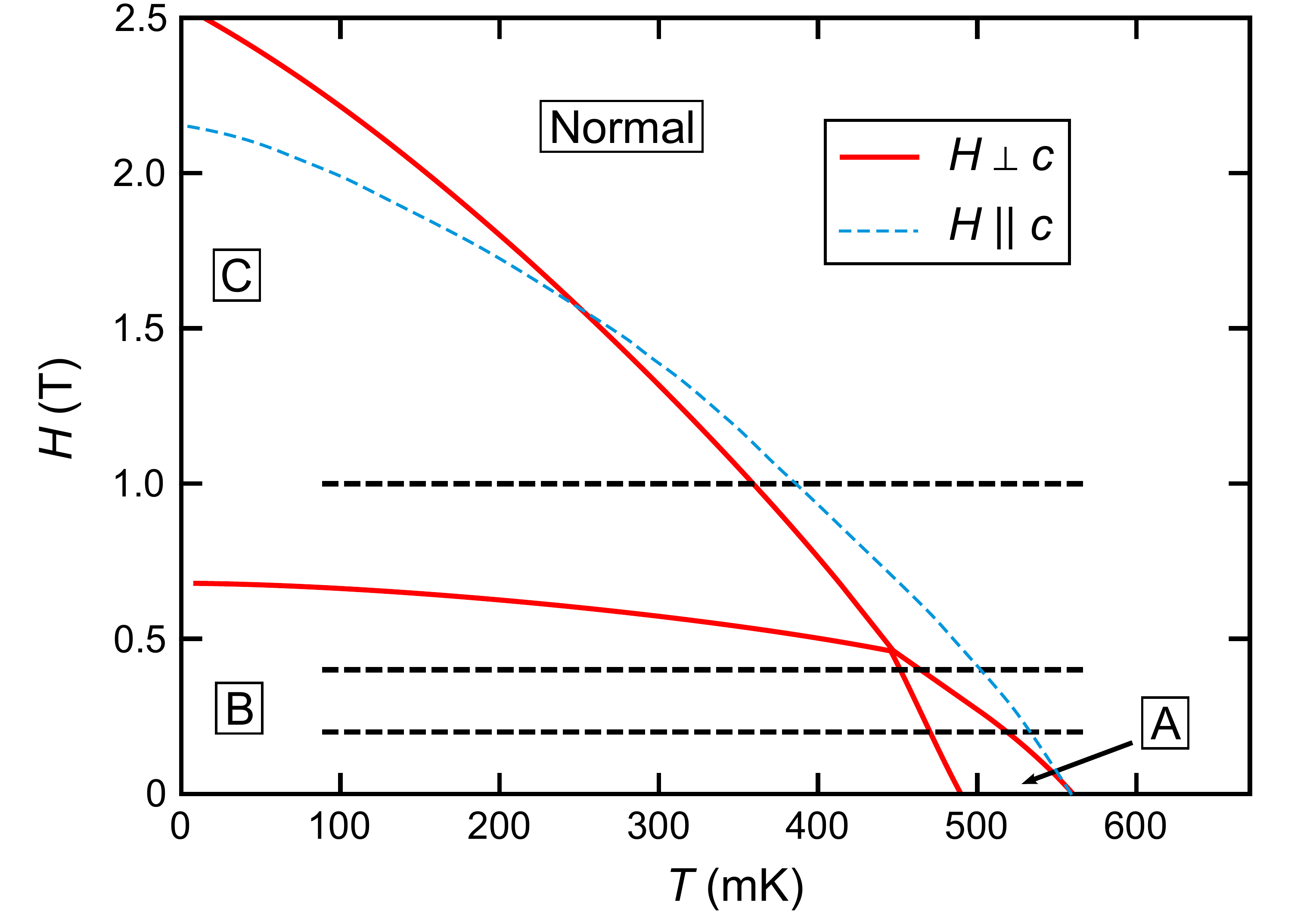}
\caption{\label{PD}(Color online) Schematic of the phase diagram of \UP\ showing the superconducting A, B, and C phases for magnetic fields perpendicular to the $c$ axis (red curves) from Adenwalla {\it et al.} (Ref. 7).  The upper critical field for fields parallel to the $c$ axis is shown as the blue dashed curve. Black dashed lines correspond to the fields at which the temperature dependence of the magnetization was measured.}
\end{figure}

Since the discovery of superconductivity in \UP\ by Stewart {\it et al.},~\cite{Stewart_PRL_1984} it has become a paradigm for unconventional superconductivity and the subject of extensive theoretical and experimental study.~\cite{Joynt_RevModPhys_2002}  The unusual properties of \UP\ include  development of a heavy-fermion state below $T$ = 20 K,~\cite{Pethick_PRL_1986} dynamic antiferromagnetism that onsets at $T_N$ = 6 K,~\cite{Lee_PRB_1993, Fomin_SSC_1996, Okuno_JPSJ_1998} and an anisotropic superconducting state with three distinct superconducting phases, two of which exist in zero applied magnetic field ~\cite{Adenwalla_PRL_1990} (Fig.\,\ref{PD}). The multiple phases provide  strong evidence for unconventional superconductivity.  Identification of  the symmetry of the order parameter requires measurement of bulk behavior in single crystals which we report here for the spin susceptibility using polarized neutron scattering.

Theoretical accounts of  many experiments that probe the orbital structure of the order parameter classify \UP\ as an odd-parity,  $f$-wave orbital state,~\cite{Sauls_AdvPhys_1994, Ohmi_JPSJ_1996, Joynt_RevModPhys_2002} with nodal structure that, in one case, has been directly observed.~\cite{Strand_Science_2010}  However, the spin character of the order parameter is not well established. In principle, polarized neutron scattering, $\mu$SR, and NMR can be used to probe the spin state, although in the first two instances there are no clear results to date and for the third, the measurements are restricted to surface regions of the sample which might be problematic in the presence of strong spin-orbit interaction. In fact, the  $^{195}$Pt NMR Knight shift, $K$,  appears to be inconsistent with measurements of the anisotropy of the upper critical field~\cite{Graf_PRB_2000} providing motivation for our work, where we measure the bulk spin susceptibility.

The NMR results indicate that the spin susceptibility in the superconducting state is unchanged from the normal state for both parallel and perpendicular orientations of the magnetic field with respect to the crystal $c$-axis.~\cite{Tou_PRL_1996, Tou_PRL_1998}  This suggests an equal-spin pairing state with the spin angular momentum always directed along the magnetic field, and is possible only if there is little or no spin-orbit coupling.~\cite{Ohmi_JPSJ_1996}  Conversely, the temperature dependence of the upper critical field, $H_{C2}$,  exhibits strong anisotropy,~\cite{Shivaram_PRL_1986} leading to the crossing of the $H_{C2}$ temperature dependence curves for the different field directions, as illustrated in Fig. \ref{PD}.  This implies Pauli limiting for only one direction of the field, $H||c$.  This temperature-dependent anisotropy requires a strong spin-orbit interaction locking the direction of zero spin projection (the direction of the so-called $d$ vector) to be parallel to the $c$-axis.~\cite{Choi_PRL_1991, Sauls_AdvPhys_1994} In this case, it is expected~\cite{Graf_PRB_2000} that the spin susceptibility, and correspondingly the Knight shift,  should decrease to zero at low temperatures for $H||c$ in contradiction to the NMR results of Tou {\it et al.}~\cite{Tou_PRL_1996, Tou_PRL_1998}

One possible explanation for this discrepancy is that the experiments  have been misinterpreted.  For example, the NMR measurements  probe only a short distance from the surface of the sample given by a London penetration depth ($\lambda_c \sim 4,000 ~\AA, \lambda_{ab} \sim 7,000 ~\AA$) where it is possible that spin-orbit scattering masks bulk behavior.~\cite{Abrikosov_ZhETF_1960, *Abrikosov_JETP_1961}  Furthermore, superconductivity near the surface of \UP\ is particularly sensitive to strain or roughness induced during crystal processing.\cite{Strand_Science_2010}  Surface sensitivity, however,  is not a concern for the measurements of the upper critical field which were performed using ultrasonic techniques.  Consequently, a true bulk probe of the magnetic susceptibility is highly desirable.  Here, we report the results of polarized neutron diffraction experiments which we have used to measure the bulk magnetization of \UP.  A similar experiment was attempted some time ago with inconclusive results.~\cite{Stassis_PRB_1986}  There have been substantial improvements in crystal quality \cite{Kycia_PRB_1998} since that time, making this a critical problem to revisit.

The large, high-quality, single crystal used in the present experiment was grown using float-zone refining in ultrahigh vacuum, followed by annealing for 6 days at $850 ^{\circ}$ C with warming and cooling each taking place over 4 days.  A 15 g portion of the crystal was cut into two parts and the residual resistance ratio was measured from three wafers cut from the middle and each end of the original crystal giving RRR$_c  = 686,  \,601, $ and $ 664$, respectively. The superconducting transition was measured resistively to be $T_c = 0.560$ K with a transition width of $\Delta T_c = 0.010$ K.  The two portions of the crystal were mounted in the $a^*-c^*$ scattering plane and attached to a copper cold finger via silver epoxy.  The two crystal sections were co-aligned to better than $0.5^{\circ}$ for all crystal axes.  The cold finger was mounted on the mixing chamber of a dilution refrigerator and cooled inside of a vertical superconducting magnet on the D3 2-axis diffractometer at the Institut Laue Langevin.  For all measurements, a neutron wavelength of 0.825 \AA was used.

The bulk magnetization was determined by measuring the flipping ratio $R$ at a nuclear Bragg reflection.  Here, $R$ is defined as the ratio of scattering cross sections for incident neutrons with spins parallel and antiparallel to the  applied magnetic field, and an arbitrary final spin state in each case.  The cross section is given by~\cite{Squires_book_1978}

\begin{equation} \label{Xsec}
\left(\frac{d\sigma}{d\Omega} \right)_{\sigma_{i}\rightarrow\sigma_{f}} \propto \,\,\,\,\,\mid \bra{\sigma_{i}} \frac{\gamma \,r_{0}}{2\mu_{B}}\vec{\sigma}\cdot \hat{\kappa}\times\{\vec{M}(\vec{\kappa})\times\hat{\kappa}\} + F_{N}(\vec{\kappa})\ket{\sigma_{f}} \mid^{2}
\end{equation} 
where $\gamma \,r_{0}$ is the classical radius of the electron multiplied by the neutron gyromagnetic ratio, $\hbar\vec{\kappa}$ is the neutron momentum transfer, $\mu_{B}$ is the Bohr magneton, $\vec{M}(\vec{\kappa})$ is the Fourier transform of the real-space magnetization induced in the sample by the applied magnetic field, and $\vec{\sigma}_{i, f}$ are the neutron's initial and final spin states.  $F_N(\vec{\kappa})$ is the nuclear structure factor for the reflection being measured.  The magnetic field was fixed in the vertical direction and the neutron spin was therefore restricted to be parallel or antiparallel to the vertical.  Only reflections in the horizontal (scattering) plane of the instrument were measured.  With these constraints, it follows from Eq. (\ref{Xsec}) that the experiment was sensitive to the component of magnetization parallel to the applied field, $M_{||}(\vec{\kappa})$.  Because the magnetic field was always along the crystal $a$-axis, we measured the magnetization in the basal plane.

In the limit that the magnetic term in Eq. (\ref{Xsec}) is much smaller than the nuclear term (\textit{i.e.} ($\gamma \,r_{0}/2\mu_{B})|M/F_N| << 1$), as is the case for the present experiment, the flipping ratio from the cross section in Eq. (\ref{Xsec}) is given by
\begin{equation} \label{R1}
R = \frac{|F_N(\vec{\kappa})-mM_{||}(\vec{\kappa})|^2}{|F_N(\vec{\kappa})+mM_{||}(\vec{\kappa})|^2}
\end{equation}
with $m = (\gamma \,r_{0}/2\mu_{B})$.  As can be seen from Eq. (\ref{R1}), changing the polarization direction effectively reverses the sign of the magnetic term.  Expanding Eq. (\ref{R1}) for small $m$ leads to
\begin{equation} \label{R2}
1-R = \frac{2\gamma r_{0}}{\mu_{B}}\frac{M_{||}(\vec{\kappa})}{F_N(\vec{\kappa})}.
\end{equation}
Thus, measuring $R$ gives the Fourier component of the total magnetization $M_{||}(\vec{\kappa})$.

\begin{figure}
\includegraphics[width=80mm]{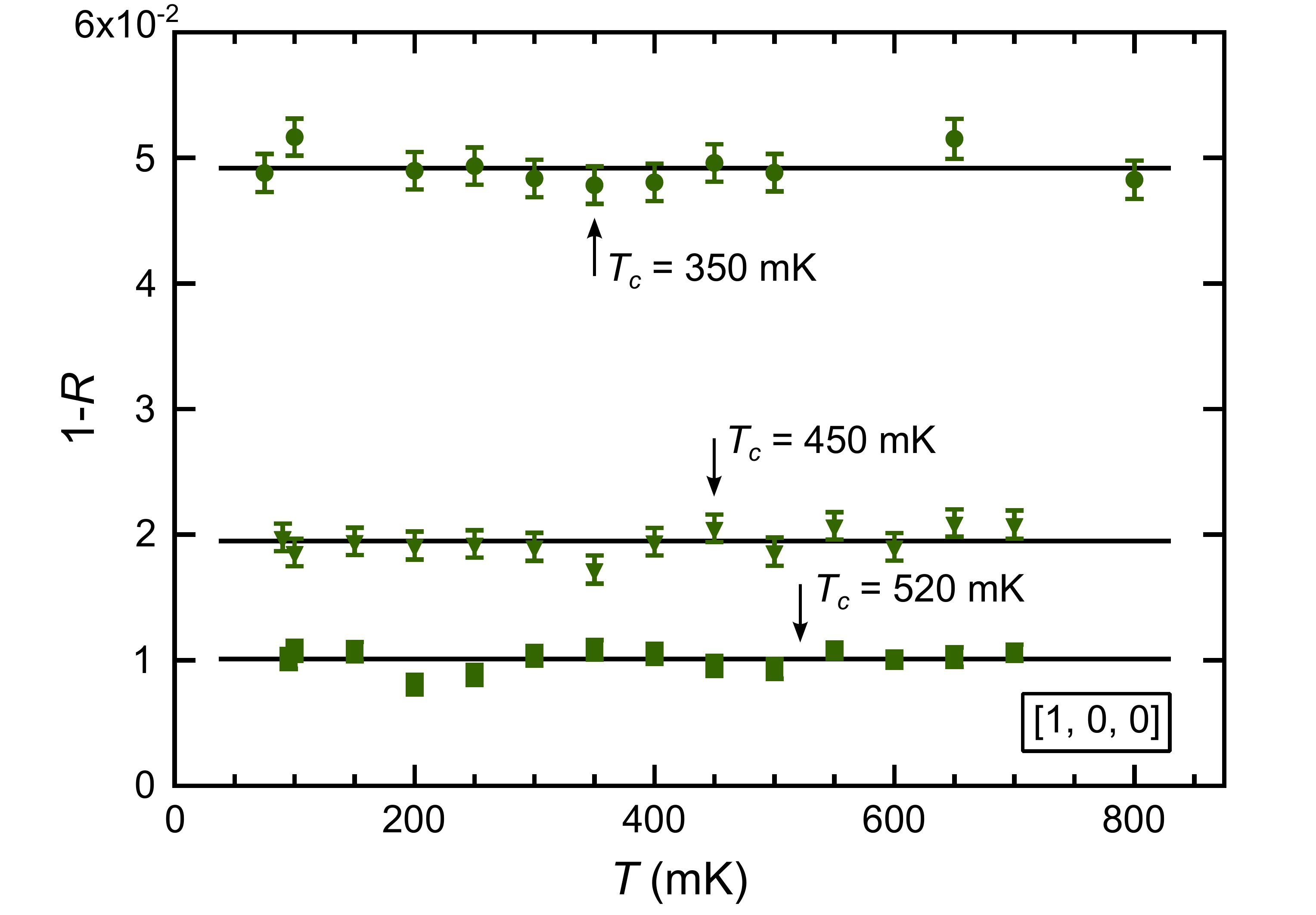}
\caption{\label{R} (Color online) The flipping ratio $R$ as a function of temperature for the nuclear Bragg reflection [1,0,0] at three fields: 0.2 T (squares), 0.4 T (triangles), and 1.0 T (circles).  Black lines show the average value of 1--$R$ at each field.}  
\end{figure}

Figure \ref{R} shows ($1-R$) at the [1,0,0] nuclear Bragg reflection for three different applied fields (note that all nuclear Bragg reflections are indexed using $a^*$ = 1.27 \AA$^{-1}$.)  The transition temperature, $T_c$\,,  for each field is indicated by a black arrow.  The data presented in Fig.\,\ref{R} for $H = 0.2, 0.4$, and 1.0 T  intercept the three superconducting phases as shown in Fig.\,\ref{PD} by the dashed lines.  In each case $1-R$ remains constant indicating that the induced magnetization does not change with temperature across the superconducting-normal transition, or for transitions between the different superconducting phases.

The [1,0,0]  reflection is ideally suited for measurements of the flipping ratio, because it is a relatively weak nuclear reflection and has the smallest possible $\vec{\kappa}$ for the present scattering configuration, maximizing $1-R$ in Eq. (\ref{R2}).  Measurements of the flipping ratio at the [2,0,0] reflection in an applied field of 1.0 T are consistent with those shown in Fig.\,\ref{R}, although with worse signal-to-noise ratio as expected for a stronger reflection at larger $\vec{\kappa}$.  Data taken at the [-1,0,0] reflection at 1.0 T lie within the error bars of the results in Fig.\,\ref{R} for  [1,0,0].

\begin{figure}
\includegraphics[width=80mm]{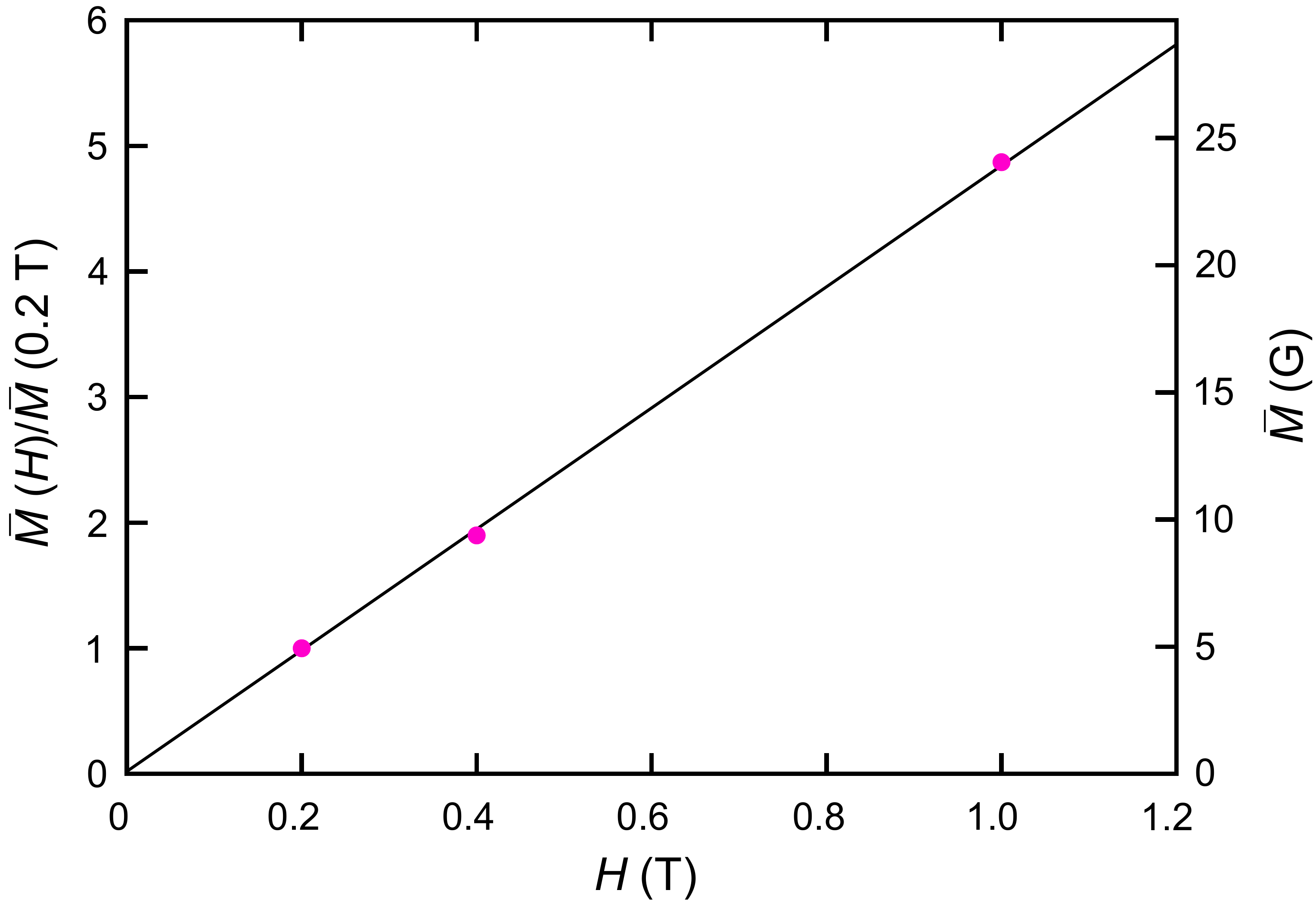}
\caption{\label{Mbar}(Color online) The average magnetization for  $T < T_c$, $\bar{M}$, as a function of applied magnetic field.  The left axis shows $\bar{M}$ normalized to the value at 0.2 T.  The right axis gives $\bar{M}$ in absolute units.  The proportionality to magnetic field indicates that the susceptibility can be taken to be $\chi = M/B$.  The zero-field $\bar{M}(H)/\bar{M}(0.2$ T$)$ intercept of the fit is 0.02.}
\end{figure}

To examine  the field dependence of the magnetization, we show its average  for $T < T_c$, $\bar{M}$, normalized to its value at 0.2 T as a function of field, Fig.\,\ref{Mbar}.  $\bar{M}$ is found to be proportional to field, and consequently the magnetic susceptibility can be taken to be $\chi = M/B$. This precise proportionality indicates that diamagnetism in the superconducting state is insignificant, since it is only weakly dependent on the magnetic field.  Additionally, we have calculated the  diamagnetism  from Ginzburg-Landau theory\cite{Brandt_PRB_2003} and find that it contributes only a small amount to the magnetization for the fields we use, consistent with the results in Fig.\,\ref{Mbar}.
 
\begin{figure}
\includegraphics[width=80mm]{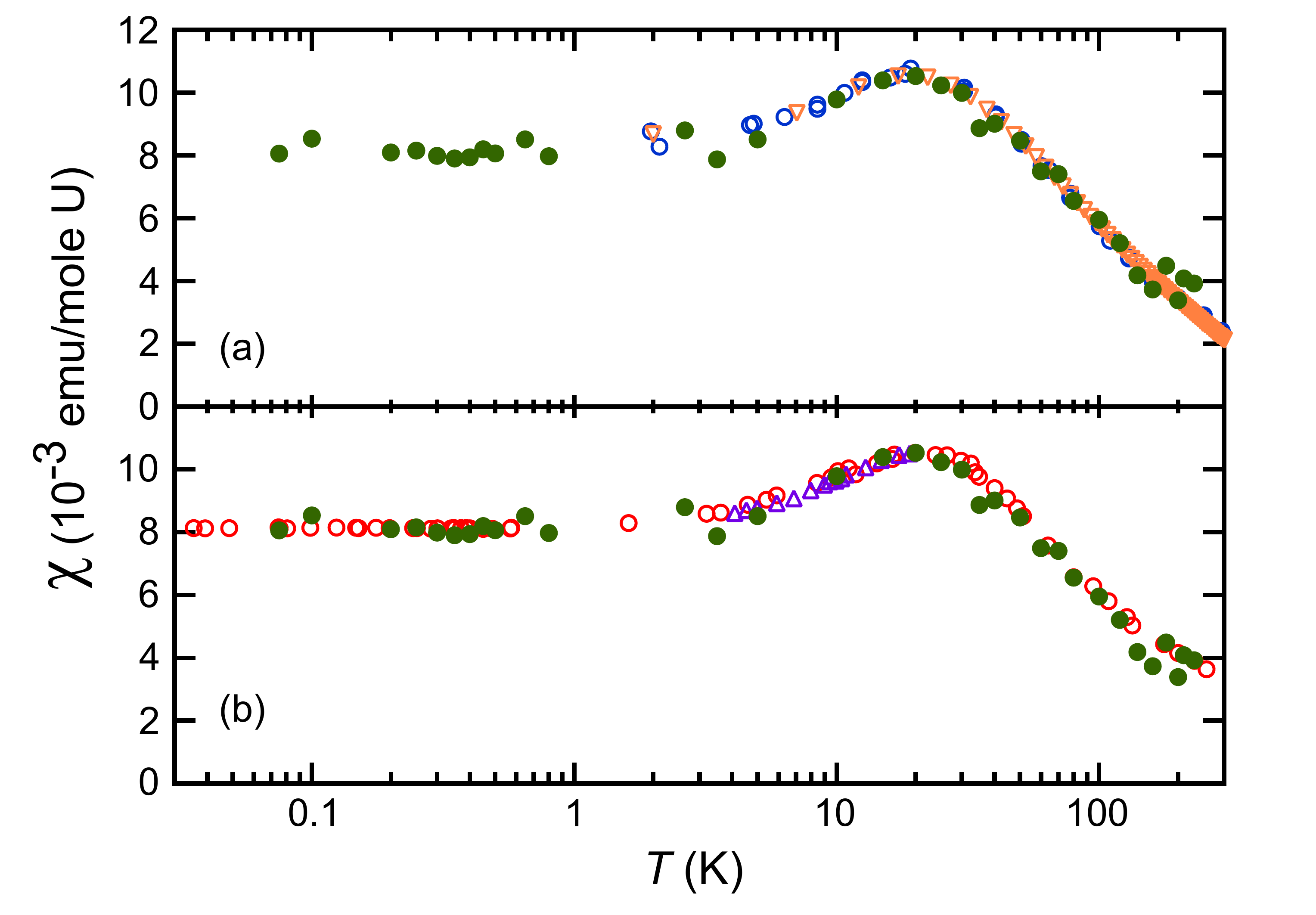}
\caption{\label{Chi}(Color online) Magnetic susceptibility measured by polarized neutron diffraction (solid green circles) multiplied by a constant factor of 2.35 (both panels) (a)  Comparison to our SQUID measurements (open orange triangles) and the susceptibility measurements from Frings \textit{et. al.} \cite{Frings_JMMM_1983}(open blue circles).  (b) Comparison to the susceptibility calculated from the Knight shift measurements as described in the text.  Open red circles are from Refs.~\onlinecite{Tou_PRL_1996, Kohori_JMMM_1990}. Open purple triangles are from Ref.~\onlinecite{Lee_PRB_1993}.  Statistical error for the susceptibility measured by neutron scattering is approximately the size of the green circles.}
\end{figure}

The polarized neutron scattering technique used here was validated in earlier applications through measurement of the temperature dependence of the magnetization of an $s$-wave superconductor, typically V$_3$Si, following the original study of this type by Shull and Wedgewood.~\cite{Shull_PRL_1966, Duffy_PRL_2000} However, \UP\  has a well-established temperature dependence of its normal-state  susceptibility that provides a convenient means to check for consistency without changing samples.  For $H||a$ there is a peak in the temperature dependence of $\chi$ at $T$ = 20 K associated with the formation of the heavy-fermion state.\cite{Joynt_RevModPhys_2002}   In addition to the low-temperature measurements shown in Figs.\,\ref{R} and \ref{Mbar}, the temperature dependence of the flipping ratio of the [1,0,0] reflection at 1.0 T has been measured from 2  to 230 K and was compared directly with measurements using a SQUID magnetometer on a small crystal of comparable quality to that of our neutron sample (also grown at Northwestern University).  Figure \ref{Chi}(a) shows the complete  temperature dependence of the susceptibility from neutron scattering at 1.0 T along with our SQUID measurements and earlier work by Frings \textit{et. al.} \cite{Frings_JMMM_1983}  The absolute value of the susceptibility was calculated from the flipping ratio using Eq. (\ref{R2}) and depends on only one parameter that is not a physical constant --- the structure factor $F_N(\vec{\kappa})$, which we calculate from the crystal structure.  We find that the neutron data match the susceptibility measurements when multiplied by a factor of 2.35.  This discrepancy is attributed to a substantial extinction correction due to the large size of our sample and $\sim 5\%$ depolarization of the incident neutron beam.

Our results can also be compared to measurements of the $^{195}$Pt Knight shift, $K$, which reflect the microscopic local field environment at the  nucleus.  $K$ is a linear function of the total magnetic susceptibility, $\chi$, and is expressed as~ \cite{Slichter_book_1990, Carter_ProgMatSci_1976} 
\begin{eqnarray} \label{Hyperfine}
K &=& K_{spin} + K_{orbit} = a_{spin}\chi_{spin} + a_{orbit}\chi_{orbit} \nonumber \\ && = a_{spin}\chi + (a_{orbit}-a_{spin})\chi_{orbit}
\end{eqnarray}
where $a_{spin}$ and $a_{orbit}$ are proportional to the respective spin and orbit hyperfine fields.  The spin hyperfine field is $H^{hf}_{spin}=a_{spin} N_{A}\mu_B$ where $N_A$ is Avagadro's number.   After noting that Knight shift contributions, other than that from electron spin, are independent of temperature, we find $H^{hf}_{spin} = -97$ kOe/$\mu_B$ from the low-temperature NMR data, similar to that reported by Lee \textit{et. al.}~\cite{Lee_PRB_1993} ($-92$ kOe/$\mu_B$) and about 12\% larger than the measurements of Tou \textit{et. al.}~\cite{Tou_PRL_1996, Tou_PRL_1998} and Kohori \textit{et. al.} \cite{Kohori_JMMM_1990} ($-85$ kOe/$\mu_B$).  In Fig.\,\ref{Chi}(b)  we show  a comparison of our polarized neutron experiment, measured in a 1.0 T field,  with the susceptibility that we calculated from the temperature dependence of the $K$ measurements of Tou \textit{et. al.}, Lee \textit{et. al.}, and Kohori \textit{et. al.} using Eq. (\ref{Hyperfine}).  Tou \textit{et. al.} \cite{Tou_PRL_1996} have estimated the size of the temperature-independent contributions with a Curie-Weiss-like fit to their $K$ data at high temperatures.  A similar fit performed to our SQUID data show in Fig.\ \ref{Chi}(a) indicates a temperature-independent susceptibly that is diamagnetic and 12.8\% of the magnitude of the average susceptibility for $T < T_c$, about a factor of 2 less than the percentage of diamagnetic Knight shift contribution found by Tou \textit{et. al.}.  The good overall agreement gives confidence that the low-temperature Knight shift results accurately represent bulk behavior of the spin susceptibility in the superconducting state.

We now discuss our results in the context of theoretical expectations for the temperature dependence of the spin susceptibility in \UP.  In a conventional superconductor, Cooper pairs form an antisymmetric spin-singlet state, commonly denoted as $\frac{1}{\sqrt{2}}\ket{\uparrow \downarrow - \downarrow \uparrow}$, leading to a decrease in the spin susceptibility for T $<$ T$_c$.   If  electron pairs form a triplet, the spins would take one of three possible arrangements, $\ket{\uparrow \uparrow}$, $\ket{\downarrow \downarrow}$, and $\frac{1}{\sqrt{2}}\ket{\uparrow \downarrow + \downarrow \uparrow}$.  For applied magnetic fields along a direction in spin-space occupied by the ``equally spin-paired'' triplet states, $\ket{\uparrow \uparrow}$ or $\ket{\downarrow \downarrow}$, no decrease in susceptibility would be expected for $T < T_c$ and H $<$ H$_{C2}$, since the field is not depairing, as is the case for superfluid $^3$He-A. \cite{Osheroff_PRL_1972} The  $K$ measurements show no (or very minimal) difference between the normal and superconducting states for all directions of magnetic field with respect to the crystal axes.  This implies that the electron spins are in an equal-spin paired state with the spin angular momentum along the direction of  the magnetic field.

For an  unconventional superconducting state, the spin part of the order parameter is best represented in terms of the $d$ vector, which points in the direction of zero spin projection and is  widely used in the description of unconventional pairing, most notably for the $^3$He triplet superfluid order parameter.\cite{Balian_PR_1963}   A strong spin-orbit interaction in \UP\ provides a mechanism for $d$ being locked to the crystal $c$ axis~\cite{Choi_PRL_1991,Sauls_AdvPhys_1994} and is required in order to account for the observed anisotropy in the upper critical field.~\cite{Shivaram_PRL_1986}  According to this theory, only the equal-spin pairing states are present in \UP\ with the spin angular momentum in the basal plane.  Their representation in terms of spin basis vectors chosen with respect to the $c$ axis, however, is $\frac{1}{\sqrt{2}}\ket{\uparrow \downarrow + \downarrow \uparrow}$ with zero spin projection in that direction.  Consequently, for strong spin-orbit interaction, a temperature independent spin susceptibility is expected for magnetic field in the basal plane and a  temperature dependent susceptibility, similar to that of a singlet superconductor is expected, for fields along the $c$ axis.

The fact that the  $K$ measurements show no decrease for any field direction in the range 0.17 T$<H<$1.6 T has been interpreted to mean that the spins are in a triplet state, but that the energy cost to rotate the $d$ vector away from the $c$ axis is quite small.\cite{Ohmi_JPSJ_1996}  This scenario is surprising as it relies on a very weak spin-orbit interaction despite expectation to the contrary owing to the heavy masses of the constituents of \UP.

In summary, our measurements of the magnetization of \UP\ with $H||a$ using polarized neutron diffraction show no change in the susceptibility upon entering the superconducting state or between the different superconducting phases.  This confirms NMR measurements of the Knight shift for this orientation.  Whether $K$ measured for $H||c$ reflects bulk behavior should be confirmed by further neutron measurements.  Although our results for $H||a$ are consistent with identification of the spin part of the superconducting order parameter as a triplet, odd-parity state, the discrepancy between the NMR measurements and measurements of the temperature dependence of $H_{C2}$ remains; a comprehensive comparison between experiment and theory for \UP\ is still incomplete.  This is also the case for other superconductors with similar properties such as Sr$_2$RuO$_4$.~\cite{MacKenzie_RMP_2003, Maeno_JPSJ_2012}

Research support  was provided by the U.S. Department of Energy, Office of Basic Energy Sciences, Division of Materials Sciences and Engineering, under Awards No.  DE-FG02-05ER46248 (Northwestern University), No.  DE-FG02- 10ER46783 (University of Notre Dame), and No. DE-FG02-05ER46202 (University of Tennessee). We acknowledge assistance with crystal growth and characterization from J. P. Davis, J. Pollanen, H. Choi, and T. M. Lippman, and we thank ILL for their hospitality and support and J. A. Sauls, V. Mitrovic, C. A. Collett, A. Zimmerman, and J. I. A. Li for helpful discussions.

\bibliography{UPt3}

\begin{thebibliography}{10}%
\makeatletter
\providecommand \@ifxundefined [1]{%
 \ifx #1\undefined \expandafter \@firstoftwo
 \else \expandafter \@secondoftwo
\fi
}%
\providecommand \@ifnum [1]{%
 \ifnum #1\expandafter \@firstoftwo
 \else \expandafter \@secondoftwo
\fi
}%
\providecommand \enquote [1]{``#1''}%
\providecommand \bibnamefont  [1]{#1}%
\providecommand \bibfnamefont [1]{#1}%
\providecommand \citenamefont [1]{#1}%
\providecommand\href[0]{\@sanitize\@href}%
\providecommand\@href[1]{\endgroup\@@startlink{#1}\endgroup\@@href}%
\providecommand\@@href[1]{#1\@@endlink}%
\providecommand \@sanitize [0]{\begingroup\catcode`\&12\catcode`\#12\relax}%
\@ifxundefined \pdfoutput {\@firstoftwo}{%
 \@ifnum{\z@=\pdfoutput}{\@firstoftwo}{\@secondoftwo}%
}{%
 \providecommand\@@startlink[1]{\leavevmode\special{html:<a href="#1">}}%
 \providecommand\@@endlink[0]{\special{html:</a>}}%
}{%
 \providecommand\@@startlink[1]{%
  \leavevmode
  \pdfstartlink
   attr{/Border[0 0 1 ]/H/I/C[0 1 1]}%
   user{/Subtype/Link/A<</Type/Action/S/URI/URI(#1)>>}%
  \relax
 }%
 \providecommand\@@endlink[0]{\pdfendlink}%
}%
\providecommand \url  [0]{\begingroup\@sanitize \@url }%
\providecommand \@url [1]{\endgroup\@href {#1}{\urlprefix}}%
\providecommand \urlprefix [0]{URL }%
\providecommand \Eprint[0]{\href }%
\@ifxundefined \urlstyle {%
  \providecommand \doi [1]{doi:\discretionary{}{}{}#1}%
}{%
  \providecommand \doi [0]{doi:\discretionary{}{}{}\begingroup
  \urlstyle{rm}\Url }%
}%
\providecommand \doibase [0]{http://dx.doi.org/}%
\providecommand \Doi[1]{\href{\doibase#1}}%
\providecommand \bibAnnote [3]{%
  \BibitemShut{#1}%
  \begin{quotation}\noindent
    \textsc{Key:}\ #2\\\textsc{Annotation:}\ #3%
  \end{quotation}%
}%
\providecommand \bibAnnoteFile [2]{%
  \IfFileExists{#2}{\bibAnnote {#1} {#2} {\input{#2}}}{}%
}%
\providecommand \typeout [0]{\immediate \write \m@ne }%
\providecommand \selectlanguage [0]{\@gobble}%
\providecommand \bibinfo [0]{\@secondoftwo}%
\providecommand \bibfield [0]{\@secondoftwo}%
\providecommand \translation [1]{[#1]}%
\providecommand \BibitemOpen[0]{}%
\providecommand \bibitemStop [0]{}%
\providecommand \bibitemNoStop [0]{.\EOS\space}%
\providecommand \EOS [0]{\spacefactor3000\relax}%
\providecommand \BibitemShut [1]{\csname bibitem#1\endcsname}%
\bibitem{Stewart_PRL_1984}%
  \BibitemOpen
  \bibfield{author}{%
  \bibinfo {author} {\bibfnamefont{G.~R.}\ \bibnamefont{Stewart}}, \bibinfo
  {author} {\bibfnamefont{Z.}~\bibnamefont{Fisk}}, \bibinfo {author}
  {\bibfnamefont{J.~O.}\ \bibnamefont{Willis}},\ and\ \bibinfo {author}
  {\bibfnamefont{J.~L.}\ \bibnamefont{Smith}},\ }%
  \bibfield{journal}{%
  \bibinfo {journal} {Physical Review Letters}\ }%
  \textbf{\bibinfo {volume} {52}},\ \bibinfo {pages} {679} (\bibinfo {year}
  {1984})%
  \bibAnnoteFile{NoStop}{Stewart_PRL_1984}%
\bibitem{Joynt_RevModPhys_2002}%
  \BibitemOpen
  \bibfield{author}{%
  \bibinfo {author} {\bibfnamefont{R.}~\bibnamefont{Joynt}}\ and\ \bibinfo
  {author} {\bibfnamefont{L.}~\bibnamefont{Taillefer}},\ }%
  \bibfield{journal}{%
  \bibinfo {journal} {Reviews of Modern Physics}\ }%
  \textbf{\bibinfo {volume} {74}},\ \bibinfo {pages} {235} (\bibinfo {year}
  {2002})%
  \bibAnnoteFile{NoStop}{Joynt_RevModPhys_2002}%
\bibitem{Pethick_PRL_1986}%
  \BibitemOpen
  \bibfield{author}{%
  \bibinfo {author} {\bibfnamefont{C.~J.}\ \bibnamefont{Pethick}}, \bibinfo
  {author} {\bibfnamefont{D.}~\bibnamefont{Pines}}, \bibinfo {author}
  {\bibfnamefont{K.~F.}\ \bibnamefont{Quader}}, \bibinfo {author}
  {\bibfnamefont{K.~S.}\ \bibnamefont{Bedell}},\ and\ \bibinfo {author}
  {\bibfnamefont{G.~E.}\ \bibnamefont{Brown}},\ }%
  \bibfield{journal}{%
  \bibinfo {journal} {Physical Review Letters}\ }%
  \textbf{\bibinfo {volume} {57}},\ \bibinfo {pages} {1955} (\bibinfo {year}
  {1986})%
  \bibAnnoteFile{NoStop}{Pethick_PRL_1986}%
\bibitem{Lee_PRB_1993}%
  \BibitemOpen
  \bibfield{author}{%
  \bibinfo {author} {\bibfnamefont{M.}~\bibnamefont{Lee}}, \bibinfo {author}
  {\bibfnamefont{G.~F.}\ \bibnamefont{Moores}}, \bibinfo {author}
  {\bibfnamefont{Y.~Q.}\ \bibnamefont{Song}}, \bibinfo {author}
  {\bibfnamefont{W.~P.}\ \bibnamefont{Halperin}}, \bibinfo {author}
  {\bibfnamefont{W.~W.}\ \bibnamefont{Kim}},\ and\ \bibinfo {author}
  {\bibfnamefont{G.~R.}\ \bibnamefont{Stewart}},\ }%
  \bibfield{journal}{%
  \bibinfo {journal} {Physical Review B}\ }%
  \textbf{\bibinfo {volume} {48}},\ \bibinfo {pages} {7392} (\bibinfo {year}
  {1993})%
  \bibAnnoteFile{NoStop}{Lee_PRB_1993}%
\bibitem{Fomin_SSC_1996}%
  \BibitemOpen
  \bibfield{author}{%
  \bibinfo {author} {\bibfnamefont{I.~A.}\ \bibnamefont{Fomin}}\ and\ \bibinfo
  {author} {\bibfnamefont{J.}~\bibnamefont{Flouquet}},\ }%
  \bibfield{journal}{%
  \bibinfo {journal} {Solid State Communications}\ }%
  \textbf{\bibinfo {volume} {98}},\ \bibinfo {pages} {795} (\bibinfo {year}
  {1996})%
  \bibAnnoteFile{NoStop}{Fomin_SSC_1996}%
\bibitem{Okuno_JPSJ_1998}%
  \BibitemOpen
  \bibfield{author}{%
  \bibinfo {author} {\bibfnamefont{Y.}~\bibnamefont{Okuno}}\ and\ \bibinfo
  {author} {\bibfnamefont{K.}~\bibnamefont{Miyake}},\ }%
  \bibfield{journal}{%
  \bibinfo {journal} {Journal of the Physical Society of Japan}\ }%
  \textbf{\bibinfo {volume} {67}},\ \bibinfo {pages} {3342} (\bibinfo {year}
  {1998})%
  \bibAnnoteFile{NoStop}{Okuno_JPSJ_1998}%
\bibitem{Adenwalla_PRL_1990}%
  \BibitemOpen
  \bibfield{author}{%
  \bibinfo {author} {\bibfnamefont{S.}~\bibnamefont{Adenwalla}}, \bibinfo
  {author} {\bibfnamefont{S.~W.}\ \bibnamefont{Lin}}, \bibinfo {author}
  {\bibfnamefont{Q.~Z.}\ \bibnamefont{Ran}}, \bibinfo {author}
  {\bibfnamefont{Z.}~\bibnamefont{Zhao}}, \bibinfo {author}
  {\bibfnamefont{J.~B.}\ \bibnamefont{Ketterson}}, \bibinfo {author}
  {\bibfnamefont{J.~A.}\ \bibnamefont{Sauls}}, \bibinfo {author}
  {\bibfnamefont{L.}~\bibnamefont{Taillefer}}, \bibinfo {author}
  {\bibfnamefont{D.~G.}\ \bibnamefont{Hinks}}, \bibinfo {author}
  {\bibfnamefont{M.}~\bibnamefont{Levy}},\ and\ \bibinfo {author}
  {\bibfnamefont{B.~K.}\ \bibnamefont{Sarma}},\ }%
  \bibfield{journal}{%
  \bibinfo {journal} {Physical Review Letters}\ }%
  \textbf{\bibinfo {volume} {65}},\ \bibinfo {pages} {2298} (\bibinfo {year}
  {1990})%
  \bibAnnoteFile{NoStop}{Adenwalla_PRL_1990}%
\bibitem{Sauls_AdvPhys_1994}%
  \BibitemOpen
  \bibfield{author}{%
  \bibinfo {author} {\bibfnamefont{J.~A.}\ \bibnamefont{Sauls}},\ }%
  \bibfield{journal}{%
  \bibinfo {journal} {Advances in Physics}\ }%
  \textbf{\bibinfo {volume} {43}},\ \bibinfo {pages} {113} (\bibinfo {year}
  {1994})%
  \bibAnnoteFile{NoStop}{Sauls_AdvPhys_1994}%
\bibitem{Ohmi_JPSJ_1996}%
  \BibitemOpen
  \bibfield{author}{%
  \bibinfo {author} {\bibfnamefont{T.}~\bibnamefont{Ohmi}}\ and\ \bibinfo
  {author} {\bibfnamefont{K.}~\bibnamefont{Machida}},\ }%
  \bibfield{journal}{%
  \bibinfo {journal} {Journal of the Physical Society of Japan}\ }%
  \textbf{\bibinfo {volume} {65}},\ \bibinfo {pages} {4018} (\bibinfo {year}
  {1996})%
  \bibAnnoteFile{NoStop}{Ohmi_JPSJ_1996}%
\bibitem{Strand_Science_2010}%
  \BibitemOpen
  \bibfield{author}{%
  \bibinfo {author} {\bibfnamefont{J.~D.}\ \bibnamefont{Strand}}, \bibinfo
  {author} {\bibfnamefont{D.~J.}\ \bibnamefont{Bahr}}, \bibinfo {author}
  {\bibfnamefont{D.~J.~V.}\ \bibnamefont{Harlingen}}, \bibinfo {author}
  {\bibfnamefont{J.~P.}\ \bibnamefont{Davis}}, \bibinfo {author}
  {\bibfnamefont{W.~J.}\ \bibnamefont{Gannon}},\ and\ \bibinfo {author}
  {\bibfnamefont{W.~P.}\ \bibnamefont{Halperin}},\ }%
  \bibfield{journal}{%
  \bibinfo {journal} {Science}\ }%
  \textbf{\bibinfo {volume} {328}},\ \bibinfo {pages} {1368} (\bibinfo {year}
  {2010})%
  \bibAnnoteFile{NoStop}{Strand_Science_2010}%
\bibitem{Graf_PRB_2000}%
  \BibitemOpen
  \bibfield{author}{%
  \bibinfo {author} {\bibfnamefont{M.~J.}\ \bibnamefont{Graf}}, \bibinfo
  {author} {\bibfnamefont{S.~K.}\ \bibnamefont{Yip}},\ and\ \bibinfo {author}
  {\bibfnamefont{J.~A.}\ \bibnamefont{Sauls}},\ }%
  \bibfield{journal}{%
  \bibinfo {journal} {Physical Review B}\ }%
  \textbf{\bibinfo {volume} {62}},\ \bibinfo {pages} {14393} (\bibinfo {year}
  {2000})%
  \bibAnnoteFile{NoStop}{Graf_PRB_2000}%
\bibitem{Tou_PRL_1996}%
  \BibitemOpen
  \bibfield{author}{%
  \bibinfo {author} {\bibfnamefont{H.}~\bibnamefont{Tou}}, \bibinfo {author}
  {\bibfnamefont{Y.}~\bibnamefont{Kitaoka}}, \bibinfo {author}
  {\bibfnamefont{K.}~\bibnamefont{Asayama}}, \bibinfo {author}
  {\bibfnamefont{N.}~\bibnamefont{Kimura}}, \bibinfo {author}
  {\bibfnamefont{Y.}~\bibnamefont{Onuki}}, \bibinfo {author}
  {\bibfnamefont{E.}~\bibnamefont{Yamamoto}},\ and\ \bibinfo {author}
  {\bibfnamefont{K.}~\bibnamefont{Maezawa}},\ }%
  \bibfield{journal}{%
  \bibinfo {journal} {Physical Review Letters}\ }%
  \textbf{\bibinfo {volume} {77}},\ \bibinfo {pages} {1374} (\bibinfo {year}
  {1996})%
  \bibAnnoteFile{NoStop}{Tou_PRL_1996}%
\bibitem{Tou_PRL_1998}%
  \BibitemOpen
  \bibfield{author}{%
  \bibinfo {author} {\bibfnamefont{H.}~\bibnamefont{Tou}}, \bibinfo {author}
  {\bibfnamefont{Y.}~\bibnamefont{Kitaoka}}, \bibinfo {author}
  {\bibfnamefont{K.}~\bibnamefont{Ishida}}, \bibinfo {author}
  {\bibfnamefont{K.}~\bibnamefont{Asayama}}, \bibinfo {author}
  {\bibfnamefont{N.}~\bibnamefont{Kimura}}, \bibinfo {author}
  {\bibfnamefont{Y.}~\bibnamefont{Onuki}}, \bibinfo {author}
  {\bibfnamefont{E.}~\bibnamefont{Yamamoto}}, \bibinfo {author}
  {\bibfnamefont{Y.}~\bibnamefont{Haga}},\ and\ \bibinfo {author}
  {\bibfnamefont{K.}~\bibnamefont{Maezawa}},\ }%
  \bibfield{journal}{%
  \bibinfo {journal} {Physical Review Letters}\ }%
  \textbf{\bibinfo {volume} {80}},\ \bibinfo {pages} {3129} (\bibinfo {year}
  {1998})%
  \bibAnnoteFile{NoStop}{Tou_PRL_1998}%
\bibitem{Shivaram_PRL_1986}%
  \BibitemOpen
  \bibfield{author}{%
  \bibinfo {author} {\bibfnamefont{B.~S.}\ \bibnamefont{Shivaram}}, \bibinfo
  {author} {\bibfnamefont{T.~F.}\ \bibnamefont{Rosenbaum}},\ and\ \bibinfo
  {author} {\bibfnamefont{D.~G.}\ \bibnamefont{Hinks}},\ }%
  \bibfield{journal}{%
  \bibinfo {journal} {Physical Review Letters}\ }%
  \textbf{\bibinfo {volume} {57}},\ \bibinfo {pages} {1259} (\bibinfo {year}
  {1986})%
  \bibAnnoteFile{NoStop}{Shivaram_PRL_1986}%
\bibitem{Choi_PRL_1991}%
  \BibitemOpen
  \bibfield{author}{%
  \bibinfo {author} {\bibfnamefont{C.~H.}\ \bibnamefont{Choi}}\ and\ \bibinfo
  {author} {\bibfnamefont{J.~A.}\ \bibnamefont{Sauls}},\ }%
  \bibfield{journal}{%
  \bibinfo {journal} {Physical Review Letters}\ }%
  \textbf{\bibinfo {volume} {66}},\ \bibinfo {pages} {484} (\bibinfo {year}
  {1991})%
  \bibAnnoteFile{NoStop}{Choi_PRL_1991}%
\bibitem{Abrikosov_ZhETF_1960}%
  \BibitemOpen
  \bibfield{author}{%
  \bibinfo {author} {\bibfnamefont{A.~A.}\ \bibnamefont{Abrikosov}}\ and\
  \bibinfo {author} {\bibfnamefont{L.~P.}\ \bibnamefont{Gor'Kov}},\ }%
  \bibfield{journal}{%
  \bibinfo {journal} {Zhurnal Eksperimentalnoi i Teoreticheskoi Fiziki}\ }%
  \textbf{\bibinfo {volume} {39}},\ \bibinfo {pages} {480} (\bibinfo {year}
  {1960})%
  \bibAnnoteFile{NoStop}{Abrikosov_ZhETF_1960}%
\bibitem{Abrikosov_JETP_1961}%
  \BibitemOpen
  \bibfield{author}{%
  \bibinfo {author} {\bibfnamefont{A.~A.}\ \bibnamefont{Abrikosov}}\ and\
  \bibinfo {author} {\bibfnamefont{L.~P.}\ \bibnamefont{Gor'Kov}},\ }%
  \bibfield{journal}{%
  \bibinfo {journal} {Sov. Phys., JETP}\ }%
  \textbf{\bibinfo {volume} {12}},\ \bibinfo {pages} {337} (\bibinfo {year}
  {1961})%
  \bibAnnoteFile{NoStop}{Abrikosov_JETP_1961}%
\bibitem{Stassis_PRB_1986}%
  \BibitemOpen
  \bibfield{author}{%
  \bibinfo {author} {\bibfnamefont{C.}~\bibnamefont{Stassis}}, \bibinfo
  {author} {\bibfnamefont{J.}~\bibnamefont{Arthur}}, \bibinfo {author}
  {\bibfnamefont{C.~F.}\ \bibnamefont{Majkrzak}}, \bibinfo {author}
  {\bibfnamefont{J.~D.}\ \bibnamefont{Axe}}, \bibinfo {author}
  {\bibfnamefont{B.}~\bibnamefont{Batlogg}}, \bibinfo {author}
  {\bibfnamefont{J.}~\bibnamefont{Remeika}}, \bibinfo {author}
  {\bibfnamefont{Z.}~\bibnamefont{Fisk}}, \bibinfo {author}
  {\bibfnamefont{J.~L.}\ \bibnamefont{Smith}},\ and\ \bibinfo {author}
  {\bibfnamefont{A.~S.}\ \bibnamefont{Edelstein}},\ }%
  \bibfield{journal}{%
  \bibinfo {journal} {Physical Review B}\ }%
  \textbf{\bibinfo {volume} {34}},\ \bibinfo {pages} {4382} (\bibinfo {year}
  {1986})%
  \bibAnnoteFile{NoStop}{Stassis_PRB_1986}%
\bibitem{Kycia_PRB_1998}%
  \BibitemOpen
  \bibfield{author}{%
  \bibinfo {author} {\bibfnamefont{J.~B.}\ \bibnamefont{Kycia}}, \bibinfo
  {author} {\bibfnamefont{J.~I.}\ \bibnamefont{Hong}}, \bibinfo {author}
  {\bibfnamefont{M.~J.}\ \bibnamefont{Graf}}, \bibinfo {author}
  {\bibfnamefont{J.~A.}\ \bibnamefont{Sauls}}, \bibinfo {author}
  {\bibfnamefont{D.~N.}\ \bibnamefont{Seidman}},\ and\ \bibinfo {author}
  {\bibfnamefont{W.~P.}\ \bibnamefont{Halperin}},\ }%
  \bibfield{journal}{%
  \bibinfo {journal} {Physical Review B}\ }%
  \textbf{\bibinfo {volume} {58}},\ \bibinfo {pages} {R603} (\bibinfo {year}
  {1998})%
  \bibAnnoteFile{NoStop}{Kycia_PRB_1998}%
\bibitem{Squires_book_1978}%
  \BibitemOpen
  \bibfield{author}{%
  \bibinfo {author} {\bibfnamefont{G.~L.}\ \bibnamefont{Squires}},\ }%
  \emph{\bibinfo {title} {Introduction to the Theory of Thermal Neutron
  Scattering}}\ (\bibinfo {publisher} {Cambridge University Press},\ \bibinfo
  {address} {England},\ \bibinfo {year} {1978})%
  \bibAnnoteFile{NoStop}{Squires_book_1978}%
\bibitem{Brandt_PRB_2003}%
  \BibitemOpen
  \bibfield{author}{%
  \bibinfo {author} {\bibfnamefont{E.~H.}\ \bibnamefont{Brandt}},\ }%
  \bibfield{journal}{%
  \bibinfo {journal} {Physical Review B}\ }%
  \textbf{\bibinfo {volume} {68}},\ \bibinfo {pages} {054506} (\bibinfo {year}
  {2003})%
  \bibAnnoteFile{NoStop}{Brandt_PRB_2003}%
\bibitem{Shull_PRL_1966}%
  \BibitemOpen
  \bibfield{author}{%
  \bibinfo {author} {\bibfnamefont{C.~G.}\ \bibnamefont{Shull}}\ and\ \bibinfo
  {author} {\bibfnamefont{F.~A.}\ \bibnamefont{Wedgwood}},\ }%
  \bibfield{journal}{%
  \bibinfo {journal} {Physical Review Letters}\ }%
  \textbf{\bibinfo {volume} {16}},\ \bibinfo {pages} {513} (\bibinfo {year}
  {1966})%
  \bibAnnoteFile{NoStop}{Shull_PRL_1966}%
\bibitem{Duffy_PRL_2000}%
  \BibitemOpen
  \bibfield{author}{%
  \bibinfo {author} {\bibfnamefont{J.~A.}\ \bibnamefont{Duffy}}, \bibinfo
  {author} {\bibfnamefont{S.~M.}\ \bibnamefont{Hayden}}, \bibinfo {author}
  {\bibfnamefont{Y.}~\bibnamefont{Maeno}}, \bibinfo {author}
  {\bibfnamefont{Z.}~\bibnamefont{Mao}}, \bibinfo {author}
  {\bibfnamefont{J.}~\bibnamefont{Kulda}},\ and\ \bibinfo {author}
  {\bibfnamefont{G.~J.}\ \bibnamefont{McIntyre}},\ }%
  \bibfield{journal}{%
  \bibinfo {journal} {Physical Review Letters}\ }%
  \textbf{\bibinfo {volume} {85}},\ \bibinfo {pages} {5412} (\bibinfo {year}
  {2000})%
  \bibAnnoteFile{NoStop}{Duffy_PRL_2000}%
\bibitem{Frings_JMMM_1983}%
  \BibitemOpen
  \bibfield{author}{%
  \bibinfo {author} {\bibfnamefont{P.~H.}\ \bibnamefont{Frings}}, \bibinfo
  {author} {\bibfnamefont{J.~J.~M.}\ \bibnamefont{Franse}}, \bibinfo {author}
  {\bibfnamefont{F.~R.}\ \bibnamefont{de~Boer}},\ and\ \bibinfo {author}
  {\bibfnamefont{A.}~\bibnamefont{Menovsky}},\ }%
  \bibfield{journal}{%
  \bibinfo {journal} {Journal of Magnetism and Magnetic Materials}\ }%
  \textbf{\bibinfo {volume} {31-34}},\ \bibinfo {pages} {240} (\bibinfo {year}
  {1983})%
  \bibAnnoteFile{NoStop}{Frings_JMMM_1983}%
\bibitem{Slichter_book_1990}%
  \BibitemOpen
  \bibfield{author}{%
  \bibinfo {author} {\bibfnamefont{C.~P.}\ \bibnamefont{Slichter}},\ }%
  \emph{\bibinfo {title} {Principles of Magnetic Resonance}},\ \bibinfo
  {edition} {3rd}\ ed.\ (\bibinfo {publisher} {Springer-Verlag},\ \bibinfo
  {address} {Berlin},\ \bibinfo {year} {1990})%
  \bibAnnoteFile{NoStop}{Slichter_book_1990}%
\bibitem{Carter_ProgMatSci_1976}%
  \BibitemOpen
  \bibfield{author}{%
  \bibinfo {author} {\bibfnamefont{G.~C.}\ \bibnamefont{Carter}}, \bibinfo
  {author} {\bibfnamefont{L.~H.}\ \bibnamefont{Bennett}},\ and\ \bibinfo
  {author} {\bibfnamefont{D.~J.}\ \bibnamefont{Kahan}},\ }%
  \bibfield{journal}{%
  \bibinfo {journal} {Progress in Materials Science}\ }%
  \textbf{\bibinfo {volume} {20}},\ \bibinfo {pages} {1} (\bibinfo {year}
  {1976})%
  \bibAnnoteFile{NoStop}{Carter_ProgMatSci_1976}%
\bibitem{Kohori_JMMM_1990}%
  \BibitemOpen
  \bibfield{author}{%
  \bibinfo {author} {\bibfnamefont{Y.}~\bibnamefont{Kohori}}, \bibinfo {author}
  {\bibfnamefont{M.}~\bibnamefont{Kyogaku}}, \bibinfo {author}
  {\bibfnamefont{T.}~\bibnamefont{Kohara}}, \bibinfo {author}
  {\bibfnamefont{K.}~\bibnamefont{Asayama}}, \bibinfo {author}
  {\bibfnamefont{H.}~\bibnamefont{Amitsuka}},\ and\ \bibinfo {author}
  {\bibfnamefont{Y.}~\bibnamefont{Miyako}},\ }%
  \bibfield{journal}{%
  \bibinfo {journal} {Journal of Magnetism and Magnetic Materials}\ }%
  \textbf{\bibinfo {volume} {90-91}},\ \bibinfo {pages} {510} (\bibinfo {year}
  {1990})%
  \bibAnnoteFile{NoStop}{Kohori_JMMM_1990}%
\bibitem{Osheroff_PRL_1972}%
  \BibitemOpen
  \bibfield{author}{%
  \bibinfo {author} {\bibfnamefont{D.~D.}\ \bibnamefont{Osheroff}}, \bibinfo
  {author} {\bibfnamefont{W.~J.}\ \bibnamefont{Gully}}, \bibinfo {author}
  {\bibfnamefont{R.~C.}\ \bibnamefont{Richardson}},\ and\ \bibinfo {author}
  {\bibfnamefont{D.~M.}\ \bibnamefont{Lee}},\ }%
  \bibfield{journal}{%
  \bibinfo {journal} {Physical Review Letters}\ }%
  \textbf{\bibinfo {volume} {29}},\ \bibinfo {pages} {920} (\bibinfo {year}
  {1972})%
  \bibAnnoteFile{NoStop}{Osheroff_PRL_1972}%
\bibitem{Balian_PR_1963}%
  \BibitemOpen
  \bibfield{author}{%
  \bibinfo {author} {\bibfnamefont{R.}~\bibnamefont{Balian}}\ and\ \bibinfo
  {author} {\bibfnamefont{N.~R.}\ \bibnamefont{Werthamer}},\ }%
  \bibfield{journal}{%
  \bibinfo {journal} {Physical Review}\ }%
  \textbf{\bibinfo {volume} {131}},\ \bibinfo {pages} {1553} (\bibinfo {year}
  {1963})%
  \bibAnnoteFile{NoStop}{Balian_PR_1963}%
\bibitem{MacKenzie_RMP_2003}%
  \BibitemOpen
  \bibfield{author}{%
  \bibinfo {author} {\bibfnamefont{A.~P.}\ \bibnamefont{Mackenzie}}\ and\
  \bibinfo {author} {\bibfnamefont{Y.}~\bibnamefont{Maeno}},\ }%
  \bibfield{journal}{%
  \bibinfo {journal} {Reviews of Modern Physics}\ }%
  \textbf{\bibinfo {volume} {75}},\ \bibinfo {pages} {657} (\bibinfo {year}
  {2003})%
  \bibAnnoteFile{NoStop}{MacKenzie_RMP_2003}%
\bibitem{Maeno_JPSJ_2012}%
  \BibitemOpen
  \bibfield{author}{%
  \bibinfo {author} {\bibfnamefont{Y.}~\bibnamefont{Maeno}}, \bibinfo {author}
  {\bibfnamefont{S.}~\bibnamefont{Kittaka}}, \bibinfo {author}
  {\bibfnamefont{T.}~\bibnamefont{Nomura}}, \bibinfo {author}
  {\bibfnamefont{S.}~\bibnamefont{Yonezawa}},\ and\ \bibinfo {author}
  {\bibfnamefont{K.}~\bibnamefont{Ishida}},\ }%
  \bibfield{journal}{%
  \bibinfo {journal} {Journal of the Physical Society of Japan}\ }%
  \textbf{\bibinfo {volume} {81}},\ \bibinfo {pages} {011009} (\bibinfo {year}
  {2012})%
  \bibAnnoteFile{NoStop}{Maeno_JPSJ_2012}%
\end{thebibliography}%

\end{document}